\documentclass[12pt]{article}
\usepackage[symbol*]{footmisc}
\usepackage{array}
\usepackage{amsmath}
\usepackage{amssymb}
\usepackage{amsfonts}
\usepackage{authblk}
\usepackage[dvips]{graphicx}
\usepackage{ccaption}

\numberwithin{equation}{section}

\input epsf
\captionwidth{.75\textwidth}

\renewcommand{\H}{\mathcal{H}}
\newcommand{\Lagr}{\mathcal{L}}
\newcommand{\imm}{\mathrm{i}}

\newcommand{\de}{\mathrm{d}}

\newcommand{\Mpl}{M_\mathrm{Pl}}
\newcommand{\V}{\mathcal{V}}
\newcommand{\R}{\mathcal{R}}
\newcommand{\K}{\mathcal{K}}

\newcommand{\sn}{\operatorname{sn}}

\newcommand{\dslash}{\not{\hbox{\kern-2pt $\partial$}}}
\newcommand{\pslash}{\not{\hbox{\kern-2.3pt $p$}}}
 \newtoks\nslashfraction
 \nslashfraction={.13}
 \newcommand{\nslash}[1]{\setbox0\hbox{$ #1 $}
   \setbox0\hbox to \the\nslashfraction\wd0{\hss \box0}/\box0 }

\begin{document}


\begin{flushright}
{NYU-TH-07/11/27}
\end{flushright}
\vskip 0.9cm

\centerline{\Large \bf Lorentz-Violating Massive Gravity}
\vskip 0.2cm
\centerline{\Large \bf in Curved Space}
\vspace{0.3in}
\vskip 0.7cm
\centerline{\large Luca Grisa}
\vskip 0.3cm
\centerline{\em Center for Cosmology and Particle Physics}
\centerline{\em Department of Physics, New York University, New York, NY 10003, USA}

\vskip 1.9cm

\begin{abstract}
	In models of modified gravity, extra degrees of freedom usually appear. They must be removed
	from the spectrum because they may indicate the presence of instabilities and because otherwise
	the model might not agree with observation. In the present letter, we will discuss a model that
	modifies General Relativity through the addition of a Lorentz-violating potential-like term
	for the metric. No additional propagating modes and no classical instabilities are present.
	The model departs from GR only in the extreme infrared region, and the cosmological evolution
	contains a bounce when the size of the Universe is of the same order of the added deformation.
\end{abstract}

\vspace{3cm}


\newpage
\section{Introduction}
%
The gravitational interaction is well described by General Relativity within a vast range of distance scales:
from solar system to galaxy cluster sizes.
Still questions arise on its applicability at microscopic, where quantum effects are no longer negligible,
and super-horizon distances.
At the far end of the length scale, in the deep infrared region, it is not obvious whether gravity should
behave the same as it does at much shorter distances. Rather, the discovery of (recent)
cosmic acceleration from supernovae data opens up the possibility
that gravity could be very different at these scales from what we experience every day.
It could be that the gravitational field is very slightly massive, and the fact was simply overlooked,
because the effects are na\"{\i}vely negligible at scales shorter than the characteristic Compton wavelength.
But, it was shown \cite{Boulware:1973my} that a simple mass term, quadratic in the fluctuation of the metric,
is not acceptable.
The classical Hamiltonian -- constructed in the ADM formalism -- is not bounded from below. The model is
not classically stable.

Modifying gravity in the IR regime seems to be rather non-trivially constrained. In the past, several
attempts to build classically stable models of IR modified gravity have been made.
For instance, extra-dimensions may manifest themselves only at very large distances, like
in the DGP model \cite{Dvali:2000hr}.
Or, without invoking extra-dimensions, the presence of non-local interactions could change the
gravitational field in the IR regime, as in \cite{ArkaniHamed:2002fu}. Or again, a similar
effect can be induced by Lorentz-violating terms, like in
\cite{ArkaniHamed:2003uy,Rubakov:2004eb,Blas:2007zz,Rubakov:2008nh}.

In \cite{Gabadadze:2004iv}, we considered a class of models of the latter group. We were able to construct
models that are four-dimensional and local, and have the property of being both stable in the IR and of
departing from standard gravitational interaction only at large distances. But this was only possible at
the price of explicitly breaking the Lorentz symmetry of the action, which had the effect of introducing
new phenomena, like instantaneous interactions. These effects can be within the experimental
bound by tuning the parameters, and more importantly they do not violate causality.

In the present letter, we will extend the previous discussion by looking at effective
models over a constantly curved background. GR will be deformed with a small Lorentz-violating
term.
The model is classically stable and behaves like a de~Sitter space-time at short scales.
At larger and larger distances, the departure from de~Sitter becomes more and more pronounced.

A rather generic feature of the model is the tendency to stop the cosmological evolution and to lead
towards a contracting phase. 
The effect of the deformation is opposite to that of
the cosmological constant, for, as a cosmological constant increases the expansion speed,
the deformation decreases it. After the expansion is stopped, {\it i.e.}, when the Hubble parameter
goes to zero, the Universe will go through a contracting phase, leading to a late-time (contracting)
de~Sitter phase.

This picture will be modified by quantum corrections. In particular, out of the gravitational field,
particles are produced when the Hubble parameter changes.
The effect of quantum produced particles is to create at late times a space-time singularity of
the same kind as the one of a supercritical Universe.
In the latter case, if the matter energy density is greater than a critical value,
the scale factor grows up
to a maximum size and then contracts towards a ``Big Crunch'' singularity: at some finite moment in the future
the scalar curvature diverges.
In the present case, instead, no criticality condition is present. The cosmological evolution is not
stopped by the matter density, but by the deformation we introduced, therefore, no matter how few particles
are produced by quantum effects, a singularity will always be reached.

The model, which is classically stable, is unstable under quantum correction. This instability is milder,
in fact its time scale can be made parametrically much greater than the present age of our Universe.
\newline
\newline
The present letter is organized as follows: we will firstly discuss the general properties of the
gravitational field in models, where a Lorentz-violating interaction for the metric is added,
explicitly showing the absence of additional degrees of freedom and the presence of instantaneous
interaction.

We will then study the cosmological solution of the (deformed) Einstein equation for two specific
choices of the Lorentz-violating interaction.
One will give rise to a bouncing Universe, while the other to a cyclic Universe of parametrically long
period. In both cases, we will consider the quantum production that occurs at the bounce, and its effect
on the cosmological evolution.

The cyclic model will then be discussed in the context of inflation. We will find that the parameters of the model
are constrained by the usual cosmological bounds. In particular, they have to be exponentially smaller
than the Hubble scale during inflation for matching density perturbation with the values observed in the CMB.

Finally, we will see the effects of the present class of deformations when in conjunction with the
massive term studied in our previous work \cite{Gabadadze:2004iv}.

\section{General Overview}
\label{gen.overview}
From a particle physicist's point of view, gravity is the interaction that emerges upon gauging the Lorentz
symmetry of Special Relativity. As for any gauge theory, it should be possible to describe the low energy
effective theory, which arises from the breaking of a part (or the whole) of the gauge symmetry as a result
of some high energy dynamics.

In gravity though, unexpected constraints arise for the low energy theory.
It was noted by Boulware and Deser \cite{Boulware:1973my} that, have the
classical non-linearities of the gravitational self-interaction taken into account, the Hamiltonian would
generally not be bounded in models of massive gravity. Thus the Boulware-Deser (BD) instability appears.

The origin of the instability -- and therefore how to render stable a model of modified gravity as originally
described in \cite{Gabadadze:2004iv} -- can be easily understood in the ADM formalism \cite{Arnowitt:1962hi},
that is the Hamiltonian formalism for gravity.

Let us construct the GR Hamiltonian.
By foliating the space-time with hypersurfaces $\Sigma_t$ for a time variable
$t$, we can replace the four-dimensional metric with the following three-dimensional variables
\begin{equation}
	\gamma_{ij}\equiv g_{ij}\,,\quad N\equiv(-^{(4)}g^{00})^{-1/2}\quad\mbox{and}\quad
	N_i\equiv\phantom{.}^{(4)}g_{0i}\,.
	\label{3D.vars}
\end{equation}
$N$ is known as the lapse function, and $N_i$ as the shift function; $\gamma_{ij}$ is the induced metric
on $\Sigma_t$. In term of these variables, we can write the four-dimensional ones as
\begin{eqnarray}
	\sqrt{-^{(4)}g}&\equiv& N\sqrt{\gamma}\,,\\
	\phantom{.}^{(4)}\R&\equiv&\phantom{.}^{(3)}\R+\K_{ij}\K^{ij}-\K^2\,,
	\label{3D.defs}
\end{eqnarray}
where $\K_{ij}$ is the extrinsic curvature on $\Sigma_t$, defined as
\begin{equation}
	\K_{ij}\equiv\frac{1}{2N}\left[ \dot\gamma_{ij}-\nabla_iN_j-\nabla_jN_i \right]\,.
	\label{K.extrcurv}
\end{equation}
The canonical momentum $\pi^{ij}\equiv\delta\Lagr/\delta\dot\gamma_{ij}$ is related to $\K_{ij}$ by the
relation
\begin{equation}
	\pi^{ij}=\sqrt{\gamma}\left[ \K^{ij}-\K\gamma^{ij} \right]\,.
	\label{pi.canmom}
\end{equation}
We now have all the ingredients to write the Hamiltonian for the Einstein-Hilbert Lagrangian
\begin{equation}
	\Lagr=\sqrt{g}\R\quad\rightarrow\quad\H\equiv\pi^{ij}\dot\gamma_{ij}-
	\Lagr|_{\dot\gamma_{ij}\mapsto\pi_{ij}}
	=\sqrt{\gamma}\left[ NR^0+N_iR^i \right]\,,
	\label{GR.H}
\end{equation}
where
\begin{eqnarray}
	R^0&\equiv&-^{(3)}\R+\gamma^{-1}\left( \pi_{ij}\pi^{ij}-\frac{1}{2}\pi^2 \right)\,,\nonumber\\
	R^i&\equiv&-2D_j(\gamma^{-1/2}\pi^{ij})\,,
	\label{R0.Ri}
\end{eqnarray}
and $D_j$ is the covariant derivative defined with respect to $\gamma_{ij}$.\\
Both $N$ and $N_i$ appear linearly in the Hamiltonian, thus they are Lagrange multipliers. The variation
with respect to them leads to the constraints -- $R^0=0$ and $R^i=0$ -- on the propagating degrees of
freedom. The Hamiltonian is exactly zero on the surface of the constraints, hence the theory is
trivially stable: the energy density of the system is bounded from below.

We will show the emergence of the BD instability for deformed
Einstein-Hilbert actions. For sake of definiteness, we shall consider the
Pauli-Fierz (PF) model \cite{Fierz:1939ix}.

The PF term is the most generic deformation
that is quadratic in the fluctuation of the metric over a particular background, and it is Lorentz-symmetric.
It describes a mass for the gravitational field
\begin{equation}
	-\frac{1}{2}m_{\mathrm{PF}}^2[h_{\mu\nu}^2-({h^\mu_{\phantom.\mu}})^2]=
	-\frac{1}{2}m_{\mathrm{PF}}^2[h_{ij}^2-h^2-2N_i^2+2h(1-N^2-N_i^2)]\,.
	\label{PF.mass}
\end{equation}
In the equality, the field $h_{\mu\nu}$ is expressed in terms of the three-dimensional variables
\eqref{3D.vars}. The tensor field is defined as $h_{\mu\nu}\equiv g_{\mu\nu}-\hat{g}_{\mu\nu}$ over
a particular background metric $\hat{g}_{\mu\nu}$. The indices are contracted using $\hat{g}^{\mu\nu}$,
{\it i.e.}, $h^\mu_{\phantom.\mu}\equiv\hat{g}^{\mu\nu}h_{\mu\nu}$, $h\equiv \hat{g}^{ij}h_{ij}$, and so on.

It is evident that the lapse and the shift functions cease to be Lagrange multipliers, and the variations
with respect to them lead to algebraic equations for them, rather than constraints on the
propagating degrees of freedom, as in GR. This is hardly unexpected. A massive field is known,
from the Lorentz group representation, to propagate a number of degrees of freedom different
from that of a massless field.

But, if we now study closely the equations
\begin{eqnarray}
	N&=&\frac{\sqrt{\gamma}R^0}{2m_{\mathrm{PF}}^2h}\,,\\
	N^i&=&\frac{1}{2m_{\mathrm{PF}}^2}(\hat{g}^{ij}-h\gamma^{ij})^{-1}R^j\,,
	\label{PF.NNi}
\end{eqnarray}
we notice that the Hamiltonian, after substituting the above values for $N$ and $N_i$,
\begin{equation}
	\H=\frac{1}{4m_{\mathrm{PF}}^2}\left[ \frac{(\sqrt{\gamma}R^0)^2}{h}
	+\gamma R^i(\hat{g}^{ij}-h\gamma^{ij})^{-1}R^j \right]
	+\frac{1}{2}m_{\mathrm{PF}}^2(h_{ij}^2-h^2+2h)\,,
	\label{PF.H}
\end{equation}
is unbounded, as it is readily seen by considering the limit
$h\rightarrow0^-$, while keeping $\sqrt{\gamma}R^0$ and $R^i=0$ fixed.

It appears that models with $N^2$-terms -- such as PF -- are generally unstable. An easy way-out is obviously
to consider more general classes of deformations, in particular the ones linear in $N$. It should be noted
though, that the $N^2$-term in the PF Hamiltonian \eqref{PF.mass} comes from the time
component of the tensor $h_{\mu\nu}$. From the definition of the lapse function \eqref{3D.vars}, it follows
that $h_{00}=g_{00}-\hat{g}_{00}\sim N^2$. Thus, removing such a term would lead to an explicit breaking
of the Lorentz symmetry in the action. For a detailed discussion of a PF-like model with
such a property we remind to our previous letter \cite{Gabadadze:2004iv}.

The class of models we would like to discuss in the present work has the following Hamiltonian
\begin{equation}
	\H=\sqrt{\gamma}\left[ NR^0+N_iR^i+2\Lambda N-2m^2Nf(\sqrt{\gamma}) \right]\,,
	\label{H.deform}
\end{equation}
where $f(\sqrt\gamma)$ is some function of the determinant of the spatial metric $\gamma_{ij}$
as in \eqref{3D.vars}, and we have assumed the presence of a cosmological constant $\Lambda$.
The deformation we added is modelled to be linear in $N$, thus the lapse function is still a
Lagrange multiplier. As in GR, the Hamiltonian is exactly zero on the solution of the constraints.
The model is therefore (classically) stable.

This model, as well as the ones presented in \cite{Gabadadze:2004iv}, should be thought of as an effective
low energy theory.
Like in the Higgs mechanism, the gauge symmetry is broken at low energy, hence we are
assuming the presence of some UV-physics that spontaneously breaks the Lorentz symmetry of the action,
like in the recent models
\cite{ArkaniHamed:2003uy},\cite{Ganor:2006ub}-\nocite{Ganor:2007qh,Dubovsky:2006vk,Dubovsky:2004sg,Bluhm:2007bd,Bluhm:2007gs}\cite{Bluhm:2008rf}.

%
The introduction of a deformation could, in principle, lead to the propagation of more degrees of freedom,
some of which may develop into instabilities for the theory. We will show that only a tranverse-traceless
tensor mode is propagating. The explicit breaking of the Lorentz symmetry will instead show up as
instanteneous interactions, as first noticed in \cite{Gabadadze:2004iv}.
To explicitly study the degrees of freedom, we turn now to the Lagrangian formalism.

The Lagrangian can be found by performing a Legendre transformation on \eqref{H.deform}
\begin{equation}
	\Lagr=\sqrt{-g}\left[ \R-2\Lambda+2m^2f(\sqrt{\gamma}) \right]\,,
	\label{L.deform}
\end{equation}
and it should be noted that we are forced to keep a somewhat mixed formalism. The deformation is
written in terms of the determinant of $\gamma_{ij}$, hence retaining in part the notion of the
three-dimensional variables \eqref{3D.vars} used in the ADM formalism.

The equations of motion are
\begin{equation}
	G_{\mu\nu}+\left[ \Lambda-m^2f(\sqrt{\gamma})
	\left( 1+\sqrt{\gamma}\frac{f'(\sqrt{\gamma})}{f(\sqrt\gamma)} \right) \right]g_{\mu\nu}
	-m^2\frac{\sqrt{\gamma}f'(\sqrt\gamma)}{|g^{00}|}\delta^0_\mu\delta^0_\nu=0\,,
	\label{eom.deform}
\end{equation}
where $G_{\mu\nu}$ is the Einstein tensor defined as $G_{\mu\nu}\equiv\R_{\mu\nu}-1/2\,\R\,g_{\mu\nu}$,
and the last term is zero for $\mu,\nu\ne0$.

The deformed action \eqref{L.deform} is not invariant under Lorentz transformations anymore. The
determinant of $\gamma_{ij}$ will transform under $x_\mu\rightarrow\Lambda_\mu^\alpha\,x_\alpha$
as it can be explicitly checked. The introduced deformation breaks the Lorentz
invariance of the action down to the rotational group. The measure on the hypersurface $\Sigma_t$
is invariant under diffeomorphisms acting on its own world-volume.

The breaking of the Lorentz symmetry stems out from the presence of a preferred frame in the model.
In the construction of the GR Hamiltonian, a frame is chosen when picking up a particular foliation
$\Sigma_t$ of the space-time. The choice is formal in GR, where the action is invariant under
the entire group of the diffeomorphisms, but it is not in the theory at hand.

The deformation we consider is a function of the spatial part $\gamma_{ij}$ of the metric tensor only.
To define $\gamma_{ij}$, we need to choose a direction $u_\mu$ and $\gamma_{ij}$ be the induced
three-dimensional metric on the hypersurface orthogonal to $u_\mu$. The preferred frame is defined
therefore by $u_\mu$. Any transformation orthogonal to $u_\mu$ is a symmetry of the model, as briefly
stated before.

Because of this remaining invariance of the action, some components of the perturbation $h_{\mu\nu}$ over
a background metric $\hat{g}_{\mu\nu}$ are not independent to the others. For sake of simplicity, we shall consider
$\hat{g}_{\mu\nu}$ to be the Minkowski metric $\eta_{\mu\nu}$. It can be shown that the results we are
presenting do not depend on this choice.

The redundancy in the components of $h_{\mu\nu}$ can be fixed by a gauge choice, for instance
$\partial^ih_{ij}=0$. It fixes the longitudinal modes of $h_{ij}$, but not its trace. The action
is indeed not invariant under the transformation
$h^i_{\phantom.i}\rightarrow h^i_{\phantom.i}+2\partial^i\xi_i$, hence
the trace $h^i_{\phantom.i}$ can not be fixed by a gauge choice.\footnote{\'Alvarez et al. in \cite{Alvarez:2006uu}
studied a similar model, where the action is symmetric under transverse diffeomorphism (TDiff), that is
$h_{\mu\nu}\rightarrow h_{\mu\nu}+\partial_{(\mu}\xi_{\nu)}$ with $\partial_\mu\xi^\mu=0$. They found
that TDiff invariant theories contain an additional scalar field; in our case the action is invariant
under spatial TDiff and no additional propagating degrees of freedom are present.}

The gauge fixing is not the only condition we have to impose on $h_{\mu\nu}$. The requirement that the equations
of motion are covariantly conserved
\begin{equation}
	(2f'+\sqrt{\gamma}f'')\partial_\mu\sqrt{\gamma}+
	\delta^0_\mu g^{0\alpha}\partial_\alpha\frac{\sqrt{\gamma}f'}{|g^{00}|}-
	g^{\alpha\beta}(\Gamma^0_{\alpha\beta}\delta^0_\mu+\Gamma^0_{\alpha\mu}\delta^0_\beta)
	\frac{\sqrt{\gamma}f'}{|g^{00}|}=0\,,
	\label{bianchi.deform}
\end{equation}
which is known as Proca condition for massive vector fields, constraints even further the independent
components of $h_{\mu\nu}$.

For perturbations over a Minkowski background, \eqref{bianchi.deform} becomes
\begin{align}
	f_0'\partial^ih_{0i}&=0\,,
	\label{proca.deform.1}\\
	(f_0'+\frac{1}{2}f_0'')\partial_kh^i_{\phantom.i}+\frac{1}{2}f_0'\partial_kh^0_{\phantom.0}&=0\,,
	\label{proca.deform.2}
\end{align}
where $f_0'\equiv f'(\sqrt{\hat{\gamma}})$ and $f_0''\equiv f''(\sqrt{\hat{\gamma}})$.

The study can be done easily using the following decomposition, common in the study of cosmological perturbations
\cite{Brandenberger:1993zc},
\begin{displaymath}
	h_{\mu\nu}^{(s)}=
	\left( \begin{array}{cc}
		\phi		& \partial_iB \\
		\partial_jB	& \eta_{ij}\chi+\partial_i\partial_jE
	\end{array}\right)\,,\quad
	h_{\mu\nu}^{(v)}=
	\left( \begin{array}{cc}
		0		& \psi_i \\
		\psi_j		& \partial_{(i}F_{j)}
	\end{array} \right)\,,\quad
	h_{\mu\nu}^{(t)}=
	\left( \begin{array}{cc}
		0		& 0 \\
		0		& h_{ij}^{\mathrm{TT}}
	\end{array} \right)\,,
\end{displaymath}
where $h_{ij}^{\mathrm{TT}}$ is a transverse and traceless tensor, $\psi_i$, $F_j$ transverse vectors,
and the rest scalars.

The scalar $E$ and the vector $F_j$ are fixed by our choice of gauge $\partial^ih_{ij}=0$. The analogous
\eqref{proca.deform.1}, \eqref{proca.deform.2} of the Proca conditions fix instead the longitudinal component
of $h_{0i}$, {\it i.e.}, the scalar $B$, and a combination of $h^0_{\phantom.0}$ and $h^i_{\phantom.i}$,
that is the solution of \eqref{proca.deform.2}.

The independent components of $h_{\mu\nu}$ are therefore: a transverse-traceless tensor $h_{ij}^{\mathrm{TT}}$,
a transverse vector $\psi_i$, and a scalar, combination of $\phi$ and $\chi$.

Not all of these five independent components are propagating degrees of freedom.
It was noticed in \cite{Gabadadze:2004iv} that
a common feature of Lorentz-violating theories is the presence of an instantaneous interaction. An explicit study
of the equations of motion for each and all components of the perturbation $h_{\mu\nu}$ would show that both
the vector and the scalar appear without time derivatives in their equations of motion. Hence, they can not be
identified as propagating degrees of freedom, rather as an instantaneous background.

The propagating degrees of freedom of the model are equivalent to the ones of standard GR, {\it i.e.}, a
transverse-traceless tensor field. The difference is the presence of instantaneous interactions. It should
not surprise, in fact breaking the gauge invariance of the action does not allow to remove from scattering
cross sections components of the metric, like the Newtonian potential $h^0_{\phantom.0}$, which is instantaneous in
nature. Thus, they would appear not only in the exchange of virtual particles, but they would also manifest
as physical phenomena.

\section{Stopping the Cosmological Expansion}
\label{sect:stopping.exp}
Having established the properties of the model under discussion, we shall now consider more specific examples.
We choose $f(\sqrt{\gamma})\equiv\gamma^{\alpha/2}$.

We are interested in particular to IR-modifications of gravity. We would like the deformation to be
dominant only at late times during the cosmological evolution, {\it i.e.}, when $a(t)\gg1$. This can be
achieved by assuming $\alpha>0$.

The equations of motion for this particular choice of $f(\sqrt{\gamma})$ are
\begin{equation}
	G_{\mu\nu}+\left[ \Lambda-(\alpha+1)\,m^2\,\gamma^{\alpha/2} \right]g_{\mu\nu}-
	\alpha\,m^2\,\frac{\gamma^{\alpha/2}}{|g^{00}|}\,\delta_\mu^{\,0}\,\delta_\nu^{\,0}=0\,.
	\label{eom.N}
\end{equation}

On the FRW ansatz with zero spatial curvature, $\de s^2=-\de t^2+a(t)^2\de\vec x^2$, they become
\begin{equation}
	\frac{\dot a^2}{a^2}-\frac{\Lambda}{3}+\frac{m^2}{3}a^{3\alpha}=0\,.
	\label{H}
\end{equation}
the space-space component of \eqref{eom.N} is proportional to the time derivative of \eqref{H}
as it can be checked explicitly.

The additional term $a^{3\alpha}$ in the Friedman equation can be mimicked by a field with
equation of state $w=-1-\alpha$. For $\alpha>0$, this fictitious field has $w<-1$.

When $m^2>0$, the energy density of the ``field'' has a wrong sign, while if $m^2<0$ its energy density is
well-behaved. The former case will be discussed in detail in the following of the present section. The
latter instead, being of a matter field with $w<-1$, will drive the cosmological expansion at an ever
increasing acceleration rate, as opposed to de~Sitter space-time of constant acceleration rate.
Such a matter field (or deformation) will drive the Universe towards a final state sometimes called
``Big Rip'': the scale factor will diverge in a finite time.

\begin{figure}[t]
	\begin{center}
	    	\epsfxsize=4in
		\epsffile{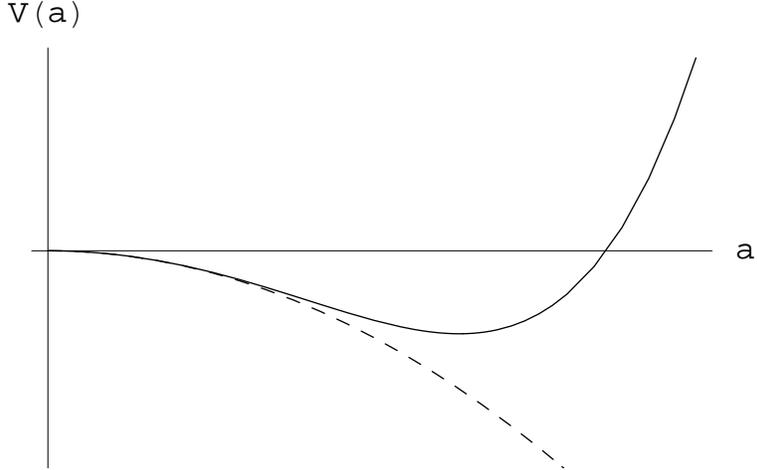}
	\end{center}
	\caption{Potential for the analogous 1-dimensional dynamics, the dashed line is the potential
	for $m=0$.}
	\label{fig:potential}
\end{figure}

Our intuition on the cosmological evolution for the $m^2>0$ case is scarcer, and this case should not be
treated as for a matter field. Let us therefore investigate the solution step by step\footnote{The study of the
cosmology dynamics in the presence of a matter field with negative energy density can be found in
\cite{Babichev:2004qp}.}.

It might be proven useful to discuss the solution of the (modified) Friedmann equation \eqref{H},
as of a classical point-particle moving in a potential
$V(a)=(-\Lambda/3+m^2/3\,a^{3\alpha})a^2$ with zero total energy.
The potential is pictured in Fig.~\ref{fig:potential}, and it should be noticed the presence of
a turning point at $a^\star=(\Lambda/m^2)^{1/3\alpha}$.

A classical point-particle moves down the potential hill from an initial position at $a=0$ until stops at $a^\star$,
because of the ``attractive force" generated by the $m^2$-term. After that position is reached, it rolls
down in opposite direction towards $a=0$, which is reached in an infinite time.

This classical analog is easily translated into the cosmological evolution of the Universe.
For small scale factor $a\sim0$, the dynamics is dominated by the cosmological
constant and the Universe is in an approximate de~Sitter phase. During this time of exponential
expansion, the ``attractive force'' of the $m^2$-term will grow in intensity, until it will
become dominant driving the cosmological evolution to a bounce at
$a(t^\star)=a^\star=(\Lambda/m^2)^{1/3\alpha}$. Then, it will contract approaching
at late times a (contracting) de~Sitter phase.

We can solve analytically the equation of motion \eqref{H}
\begin{equation}
	a(t)=\left(\frac{\Lambda}{m^2}\right)^{\frac{1}{3\alpha}}
	\left(\cosh\frac{\sqrt{3\Lambda}\,\alpha}{2}(t-t_0)\right)^{-\frac{2}{3\alpha}}\,,
	\label{sola.N}
\end{equation}
where $t_0$ is an integration constant to be fixed by imposing the initial condition $a|_{t=0}=1$:
$t_0=\frac{2}{\sqrt{3\Lambda}\alpha}\cosh^{-1}\sqrt{\frac{\Lambda}{m^2}}$.

\begin{figure}[t]
  	\centerline{\hbox{ \hspace{0.0in} 
	    \epsfxsize=2.5in
	    \epsffile{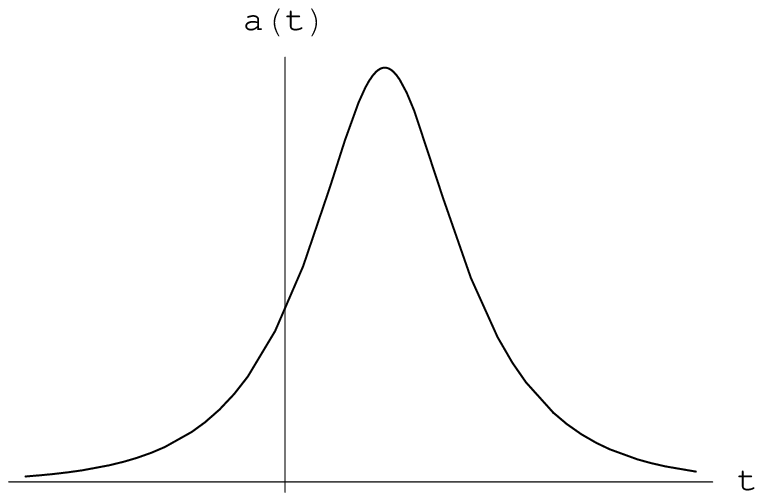}
	    \hspace{0.25in}
	    \epsfxsize=2.5in
	    \epsffile{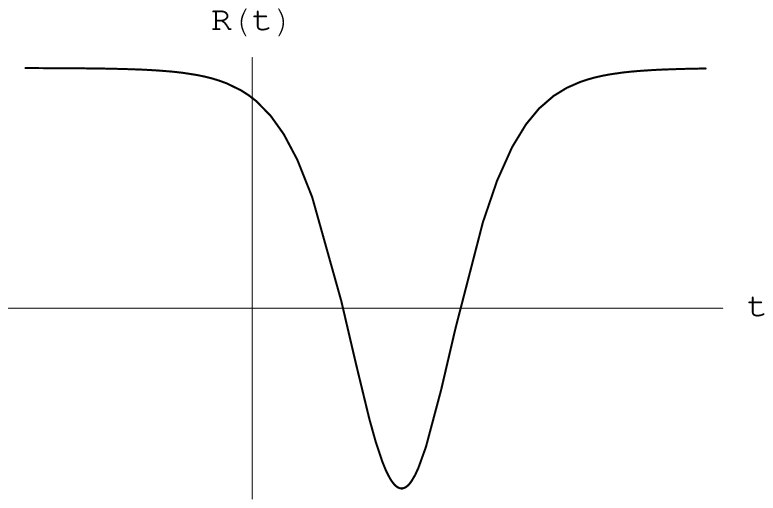}
	}}
	\caption{Plot of $a(t)$ and $\mathcal{R}(t)$.}
	\label{fig:aR}
\end{figure}

This exact solution is plotted for some positive value of $\alpha$ alongside with the Ricci curvature $\mathcal{R}(t)$ in Fig.~\ref{fig:aR}. The features previously described
are easily recognizable: at early and late time, the expansion is dominated by the cosmological constant and the Universe exponentially expands and contracts respectively.

The dynamics presented here is classical, and would be modified by quantum corrections. In particular it is known that, whenever the scalar curvature changes in time,
particles are created via a phenomenon similar to the Hawking radiation for a Black Hole.

To understand it, let us consider a basis of particle creation/annihilation operators $\{A_k^\dagger\,,A_k\}$. The Fock space is defined by determining the vacuum state
$|\Omega\rangle$, which is destroyed by all the annihilation operators, $A_k|\Omega\rangle=0$,
and then populating it by acting with the creation operators onto $|\Omega\rangle$.

The operators are constructed by canonically quantizing the fields of the particular theory at hand,
but in doing so we should assume a particular background metric. Hence,
if the metric is time-dependent, so are the operators. The vacuum $|\Omega\rangle$ is annihilated
by all $A_k$ at a given time, but not, in general, at every time. Thus as time
goes by, the vacuum state will be in a superposition of particles, for $A_k(t)|\Omega\rangle\ne0$
at a generic late time.

Since the background we found is time-dependent, we would like to study the effects quantum particle production has on the cosmological evolution.

For sake of clarity we specialize to $\alpha=1$. Following \cite{Ford:1986sy}, the energy density $\rho_\mathrm{q}$
created up to a time $\bar t$ is given by the following expression
\begin{equation}
	\rho_\mathrm{q}=-\frac{1}{32\pi^2a(\bar t)^4}\int_{-\infty}^{\bar t}\de t_1\int_{-\infty}^{\bar t}\de t_2
	\log\left|\frac{\eta_1-\eta_2}{\eta_0}\right|\,\V'(t_1)\V'(t_2)\,,
	\label{partprod}
\end{equation}
where $\V'(t)=(1-6\xi)(\dot a^2+a\,\ddot a)$ for the FRW ansatz and $\xi=1/6$ for conformally coupled fields;
$\eta$ is the conformal time defined as $d\eta=dt/a(t)$
\begin{equation}
	\eta=\eta_0\,I\left(\frac{1}{\cosh^2\frac{\sqrt{3\Lambda}}{2}(t-t_0)};-\frac{1}{3},\frac{1}{2}\right)
	\quad\mbox{with }
	\eta_0\equiv\left(\frac{m^2}{\Lambda}\right)^{1/3}
	\frac{B(-\frac{1}{3},\frac{1}{2})}{\sqrt{3\Lambda}}\,,
	\label{conftime}
\end{equation}
and $I(z;a,b)$ is the regularized beta function $I(z;a,b)=B(z;a,b)/B(a,b)$, with $B(z;a,b)$ and $B(a,b)$ being the incomplete and complete beta function respectively.

For sake of simplicity we approximate the quantum corrected dynamics as if particles were generated all at once when $\mathcal{R}(\bar t)=0$. This rough approximation will not
change the qualitative picture we will describe.

By evaluating the integral of \eqref{partprod} up to $\bar t$, we find the energy density to be
\begin{equation}
	\rho_\mathrm{q}=\frac{\mathcal{I}}{288\pi^2}\Lambda^2\sim\Lambda^2\,,
	\label{partprod.value}
\end{equation}
where $\mathcal{I}$ is the numerical result of the (adimensional) integral of \eqref{partprod}.

As it was noticed in \cite{Ford:1986sy}, $\rho_\mathrm{q}$ does not depend on the time when the de~Sitter expansion stops, but solely on the change in the scalar curvature. In our model, this
means $\rho_\mathrm{q}$ is proportional to $\Lambda$, but not to $m^2$.

After $\bar t$, the cosmological evolution will change to include the presence of matter. The analytical solution of the Friedmann equation is
\begin{equation}
	a\!=\!\left(\frac{\Lambda}{2m^2}+\sqrt{\frac{\Lambda^2}{4\mu^4}+\frac{\rho_\mathrm{q}}{\Mpl^2m^2}}\right)^{1/3}\!\!\!
	\sn^{2/3}\!\left(\frac{\sqrt\frac{3m^2\rho_\mathrm{q}}{2\Mpl^2\Lambda^2}(t-\bar t)}
	{\sqrt{1+\!\sqrt{1\!+\!\frac{4m^2\rho_\mathrm{q}}{\Mpl^2\Lambda^2}}}},
	\frac{1+\!\sqrt{1\!+\!\frac{4m^2\rho_\mathrm{q}}{\Mpl^2\Lambda^2}}}{1\!-\!\sqrt{1+\!\frac{4m^2\rho_\mathrm{q}}{\Mpl^2\Lambda^2}}}\right),
	\label{solarho.N}
\end{equation}
where $\sn(u,\mu)$ is one of the Jacobi elliptic functions.

\begin{figure}[t]
	\begin{center}
	    	\epsfxsize=4in
		\epsffile{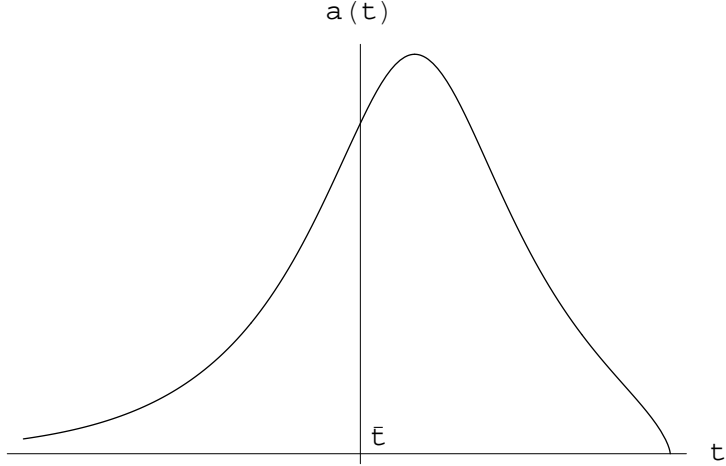}
	\end{center}
	\caption{Evolution of the scale factor considering particle creation at $\bar{t}$.}
	\label{fig:reheating}
\end{figure}

The plot of the scale factor $a(t)$ obtained by matching \eqref{sola.N} with \eqref{solarho.N} at $t=\bar t$ is shown in Fig.~\ref{fig:reheating}.

At $t\gtrsim\bar t$, the $m^2$-term is driving the cosmological evolution, in fact to have it to
stop the exponential expansion driven by $\Lambda$ -- that is to approach $\R(\bar t)=0$ --
its ``strength" has to be of the same order of the cosmological constant. Because $m^2$ is dominant,
the cosmological evolution does not change significantly from the one we described without matter.
The scale factor reaches a maximum, and then starts to contract. The dynamics starts to depart
from \eqref{sola.N} at this point.
Instead of approaching a contracting de~Sitter phase with constant curvature $\Lambda$, at a distant but
finite time in the future the Ricci curvature diverges: a ``Big Crunch" occurs.
At late times, when -- during the contracting phase -- $a(t)\ll1$, the matter density is driving the
evolution, and therefore the singularity is unavoidable.

This picture is similar to a supercritical Universe, with the difference that no criticality
condition for $\rho_\mathrm{q}$ is present. The expansion is not stopped by the (supercritical) matter density, but by
the IR modification we introduced. Thus no matter how small $\rho_\mathrm{q}$ is, the dynamics ends with a singularity at a finite time.

The model, despite being classically stable, is unstable under quantum corrections.

Moreover, even considering that the ``Big Crunch" will occur in a (parametrically) very distant future,
the exit of the de~Sitter expansion, which could be associated with
a period of inflation, coincides with an era dominated by the $m^2$-term.
Thus, the model would not follow the known cosmological evolution of our Universe.

\section{Cyclic Universe}
\label{sect:cyclic}
As we have discussed in the previous section, a ``Big Crunch" is unavoidable once quantum corrections are
taken into account. One way to avoid the singularity could be to modify once more the gravitational Lagrangian.

By studying the effect the $m^2$-term has on the cosmological evolution, we understood that the modification we introduced slows down the expansion and ultimately stops it, after which the Universe goes through a contracting phase.
If we could have a term that acts as the $m^2$-term, namely that slows down the cosmological evolution,
but is dominant only when the size of the Universe is small, we might stop the
collapse before the scale factor reaches zero, and the curvature diverges. This can be attained by having chosen the exponent $\alpha$ to be negative.

The most general (and minimal) Lagrangian is
\begin{equation}
	\Lagr=\sqrt{-g}\left[\mathcal{R}-2\Lambda+2m^2\gamma^{\alpha/2}+2k^2\gamma^{\beta/2}\right]\,,
	\label{lagr.cyclic}
\end{equation}
where $\alpha$ is positive and $\beta$ negative definite. On the FRW ansatz, the equation of motion is
\begin{equation}
	\frac{\dot a^2}{a^2}+\frac{m^2}{3}a^{3\alpha}+\frac{k^2}{3}a^{3\beta}-\frac{\Lambda}{3}=0\,.
	\label{eom.cyclic}
\end{equation}

\begin{figure}[t]
	\begin{center}
	    	\epsfxsize=4in
		\epsffile{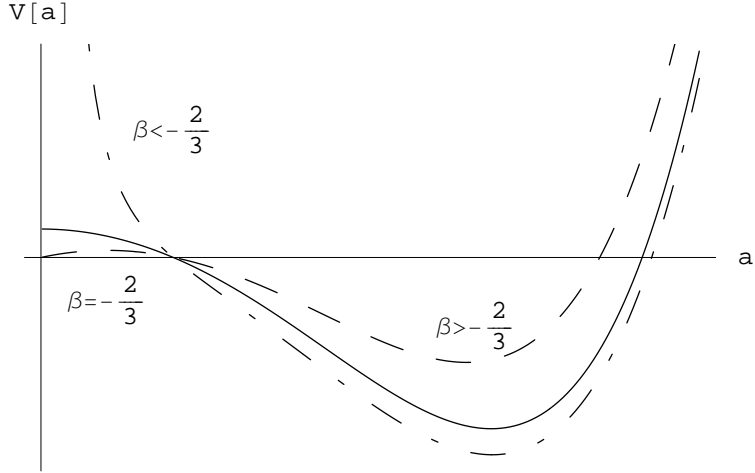}
	\end{center}
	\caption{Potential for the analogous 1-dimensional dynamics; the continuous line is for $\beta=-2/3$,
	the dashed for $\beta>-2/3$ and the dot-dashed for $\beta<-2/3$.}
	\label{fig:pot-muk}
\end{figure}

We can study the dynamics of $a(t)$ by analogy with a classical particle moving in the potential $V=(m^2a^{3\alpha}-\Lambda+k^2a^{3\beta})a^2/3$. The potential is
plotted in Fig.~\ref{fig:pot-muk}, for $\beta<-2/3$, $\beta=-2/3$ and $\beta>-2/3$. The common feature of all the plotted potentials is the presence of two turning points $a_{(\pm)}$,
between which the point-particle would oscillate back and forth.

The model is of a cyclic Universe that ``eternally" oscillates between a minimum and a maximum size. Between the two extrema, the Universe is in a de~Sitter phase, 
either contracting or expanding.

\begin{figure}[t]
	\begin{center}
	    	\epsfxsize=4in
		\epsffile{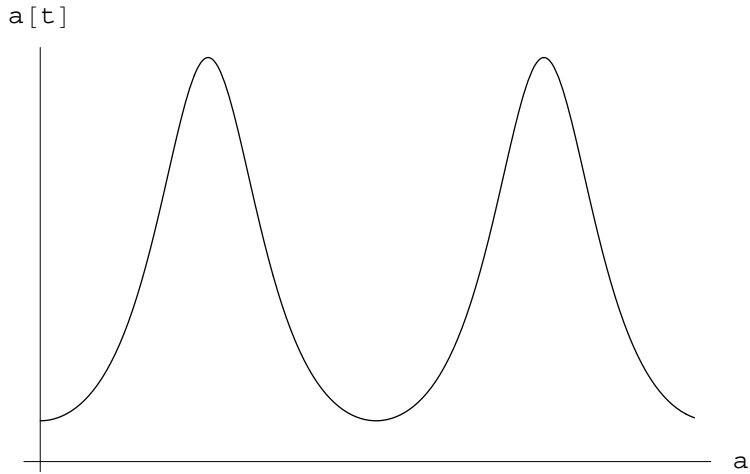}
	\end{center}
	\caption{Plot of the scale factor $a(t)$ in respect to time for the cyclic solution.}
	\label{fig:cyclic}
\end{figure}

We can study analytically the dynamics for the choice of $\alpha=2/3$ and $\beta=-2/3$; the solution,
shown in Fig.~\ref{fig:cyclic}, is
\begin{equation}
	a(t)=a_{(-)}\left[1-\sn^2\left(\imm\sqrt{\frac{a_{(+)}^2-a_{(-)}^2}{3}}m(t-t_0),
	\frac{a_{(-)}^2}{a_{(-)}^2-a_{(+)}^2}\right)\right]^{1/2}\,,
	\label{sol.cyclic}
\end{equation}
where $a_{(\pm)}\equiv(\Lambda\pm\sqrt{\Lambda^2-4k^2m^2})/2m^2$, and $\sn(u,\mu)$ is the Jacobi $\sn$ elliptic function with periodicity\footnote{more exactly, the Jacobi
$\sn(u,\mu)$ is doubly periodic in the complex plane,
that is $\sn(u+4(K(\mu)+\imm\,n\,K(1-\mu)),\mu)=\sn(u,\mu)$ where $n\in\mathbb{Q}$.
In the present case, $n$ is fixed by requiring the period to be real.}
\begin{equation}
	T=\frac{4\sqrt{3}}{(\Lambda^2-4k^2m^2)^{1/4}}
		\left[\imm K\left(\frac{a_{(-)}^2}{a_{(-)}^2-a_{(+)}^2}\right)
		+K\left(\frac{a_{(+)}^2}{a_{(-)}^2-a_{(+)}^2}\right)\right]\,,
	\label{period.cyclic}
\end{equation}
where $K(\mu)$ is the complete elliptic integral of first kind.

The presence of the $k^2$-term effectively ``screens" regions of small size. It slows and stops
the contraction up to a non-zero scale factor $a_{(-)}$, in the same way that
the $m^2$-term stops the expansion at large scale.

Na\"{\i}vely we would expect this model to be stable under quantum correction, for $k^2$-term could stop the collapse in presence of matter too. It is like the new
term creates a potential barrier at small scales, preventing the scale factor to reach zero size. The presence of matter would lower the barrier, but by tuning $k^2$ we can
avoid its disappearance.

The loop-hole in this argument is the periodicity of the model. At each cycle new matter is generated by
quantum corrections as seen in the previous section.
Eventually $\rho_\mathrm{q}$ will ``overcome" the barrier leading towards a ``Big Crunch", unless the $k^2$-deformation
is always dominant at small scale factors, no matter how big $\rho_\mathrm{q}$ is. This is obtained by taking 
$\beta<-1-w$, where $w$ is the equation of state for the particles generated through quantum effects.
For radiation $w$ is equal to $1/3$, thus if $\beta<-4/3$ a minimal size $a_{(-)}$ is always present.

In the approximation the $m^2$-deformation and cosmological constant are negligible, which is always the case for
$a(t)\ll1$, the minimal size is
\begin{equation}
	a_{(-)}\sim\left( \frac{k^2}{\rho_\mathrm{q}} \right)^{1/3|\beta+1+w|}\,.
	\label{cyclic.minsize}
\end{equation}

Obviously, because more and more matter is generated at each cycle, a time will come
when $a_{(-)}$ will be of Planck size, and therefore our semi-classical description will break down.


\section{Embedding Inflation}
In the following, we would like to depart slightly from the previous discussions, that have considered
the effects of gravity modifications in a de~Sitter space-time.

\begin{figure}[t]
	\begin{center}
	    	\epsfxsize=4in
		\epsffile{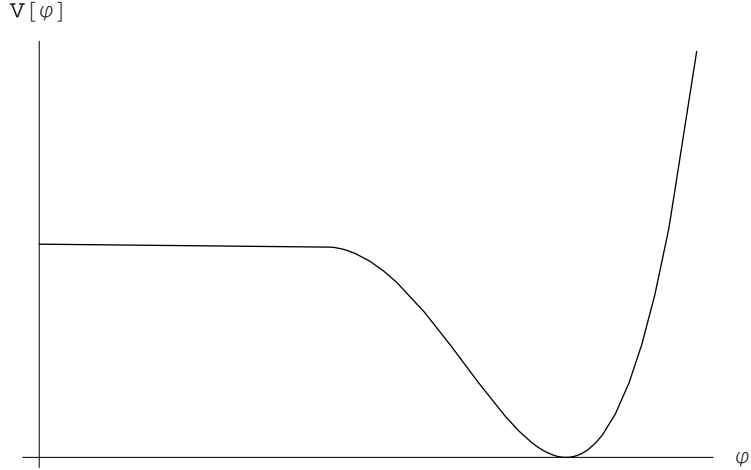}
	\end{center}
	\caption{Typical inflaton potential $V(\varphi)$.}
	\label{fig:inflaton}
\end{figure}

We will consider a generic inflationary model, and describe how the terms we introduced would effect inflation, and what kind of bounds we could have. We will, therefore,
consider modified gravity coupled to a scalar field, the inflaton.
The plot of a typical inflaton potential $V(\varphi)$ is sketched in Fig.~\ref{fig:inflaton}.

When the inflaton is atop the plateau, its energy density is dominated by the potential, namely $\dot\varphi^2\ll V(\varphi)$, and approximately constant. This plays the role
of the cosmological constant $\Lambda=V(\varphi)$. The space-time is exponentially expanding as long as the inflaton sits on the plateau.

This configuration is not stable, in fact it is energetically favorable for the scalar field to condense at the minimum of the potential, $\langle\varphi\rangle=\varphi_0$. When the
inflaton condenses on its true vacuum, inflation ends, because the energy density is no longer a non-zero constant.

Once inflation is over, the inflaton decays more or less efficiently, depending on its coupling, into the Standard Model fields, reheating the Universe. From this point on, the
cosmological evolution follows the Standard Model of Cosmology.

We can embed this general picture into our model and describe what constraints arise from cosmology and general requirements.

The Universe described by \eqref{sol.cyclic} -- solution of the Friedmann equation -- goes through
periodic exponential expansions and contractions. The requirement to have
the usual description from the inflation framework within this model demands the scalar field $\varphi$ to condense into its true vacuum during an exponential expansion.
Conversely, if it condensed during a contraction, the density perturbation originated during the inflationary period would be washed out at the unavoidable bounce.

Thus the inflaton can condense any time between $a_{(-)}$ and $a_{(+)}$, and this in turn provides a bound on the number of e-foldings $N_e$ as a function of the parameters
$m^2$ and $k^2$
\begin{equation}
	N_e\geqslant\log\frac{a_{(+)}}{a_{(-)}}=\log\frac{\Lambda+\sqrt{\Lambda^2-4k^2m^2}}{\Lambda-\sqrt{\Lambda^2-4k^2m^2}}\,.
\end{equation}

This relation can also be read as a bound on the parameter $m^2$. Assuming $m^2$ to be much smaller than $\Lambda$ and $k^2$, we find the following bound
\begin{equation}
	m^2\lesssim\frac{\Lambda^2}{k^2}\exp[-2\,N_e]\,.
	\label{mu2bound.cyclic}
\end{equation}

The modification we introduced must be exponentially smaller than the de~Sitter curvature in order
to satisfy the cosmological bound on $N_e$ as it arises from matching the
inflaton density perturbations with the anisotropies observed in the\linebreak CMB ($N_e\gtrsim60$).

Few comments should follow. Firstly, \eqref{mu2bound.cyclic} is strictly an upper bound, for, if inflation end when the $m^2$-term is dominant, that is when $a(t)\sim a_{(+)}$,
the cosmological evolution could not start in a radiation-dominated epoch, as in the Standard Model of Cosmology.

Secondly, whether in our model inflation could be eternal -- see, for instance \cite{Linde:2004kg,Guth:2007ng}.
As we have throughly discussed in the previous sections, the additional potential-like term acts
to stop the cosmological evolution bringing to an end any inflationary periods. It is not obvious how
inflation could be eternal in a model that does not allow the space time to expand indefinitely.

In the standard picture of eternal inflation, false vacuum bubbles nucleate due to quantum fluctuations of
the inflaton field. If the size of these bubbles is larger than their Hubble radius, they are causally
disconnected from the ``ambient'' space. They will independently evolve and eventually pinch off.

As long as the deformation is negligible, the previous picture applies to our model as well.
From the point of view of an observer living in one of these bubbles, they
have no way to know anything about what is happening outside the horizon. It should be natural to expect
that, once the ``ambient'' space bounces back, it keeps collapsing towards a de~Sitter contracting
phase, while the bubbles, unaware of anything happening outside their Hubble radius, expand until
they pinch off. Hence new ``baby'' Universes are generated, and they will follow the same evolution of
their ``parent'' Universe: the Universe will keep self-replicating and at any time at least one patch
will be in an inflationary regime.

Even though every bubble can expand only till reaching a finite size before bouncing back towards a contracting
phase, the total number of nucleated bubbles will be infinite, hence eternal inflation is a possible
scenario in our model.

\subsection{A Curiosity}
Something curious happens for a particular, non-small value of $km/\Lambda$.

If $\Lambda=2km$, the two turning points $a_{(\pm)}$ are equal as it follows from their definitions.
At this particular value, the solution of the Friedmann equation is of a flat space-time.

For the tuned value of $m^2$, an otherwise de~Sitter space-time turns out to be effectively flat. This solution can be nicely understood from the analogous classical point-particle
description. As we have already stressed, the cosmological constant acts as a repulsive force, whilst the $m^2$-term as an attractive one. When $\Lambda=2km$, those two
forces balance exactly leading to the allowed range of $a(t)$ to shrink down to a point. From the cosmological view point, the scale factor is time-independent and therefore
the space-time is effectively flat.

Some fine-tuning is required for this solution, therefore it would be interesting to understand how (if)
the solution is reached in a generic inflationary model.

Let us consider a linear potential $V(\varphi)=\Lambda_0-\eta^2\varphi$ with $\eta^2\ll\Lambda_0$ to satisfy the slow-roll condition. In a usual inflationary model, inflation would be
eternal, for no minimum of the potential of the scalar field is present.
But an exponential expansion always stops when the $m^2$-term becomes dominant, so it is not clear
what dynamics will follow in our model.

The scalar field follows its own equations of motion coupled to gravity
\begin{eqnarray}
	\ddot\varphi+3\frac{\dot a}{a}\dot\varphi+V'(\varphi)&=&0\,,\nonumber\\
	\frac{\dot a^2}{a^2}+m^2a^2+k^2a^{-2}-\left[ \frac{1}{2}\dot\varphi^2+V(\varphi) \right]&=&0\,,
	\label{eom.slowroll}
\end{eqnarray}
where $H=\dot a/a$, the Hubble constant, is a friction force for $\varphi$. In the slow-roll approximation,
the inflaton reaches critical but small velocity
$|\dot\varphi|\sim|V'(\varphi)/3H|=\eta^2/3H\ll V(\varphi)$.

The effective cosmological constant is $\Lambda\sim V(\varphi)=\Lambda_0-\eta^2\varphi$. Since $\varphi$ is changing in time, so is $\Lambda$, but we can assume the slow-roll
approximation to hold at any time, $\dot\varphi^2\ll V(\varphi)$.

The Hubble constant $H^2=(\dot a/a)^2$ is monotonically decreasing from its initial value $\Lambda_0$.
The lower limit is at $\Lambda\sim2km$, where $\dot\varphi$ diverges. Thus, at this point,
the inflaton is in a fast-roll regime with the kinetic energy dominating over the potential: $\dot\varphi^2\gg V(\varphi)$.

We can drop the potential from the equations of motion \eqref{eom.slowroll}
\begin{eqnarray}
	\ddot\varphi+3\frac{\dot a}{a}\dot\varphi&=&0\,,\label{phi.fastroll}\\
	\frac{\dot a^2}{a^2}+m^2a^2+k^2a^{-2}-\frac{1}{2}\dot\varphi^2&=&0\,.\label{a.fastroll}
\end{eqnarray}

The dominant term is the kinetic energy -- $\dot\varphi^2=\dot\varphi^2_0a^{-6}$ from \eqref{phi.fastroll} --
that drives the expansion until, at very late times, the $m^2$-term becomes dominant and leads to a contracting phase.

Because $\dot\varphi_0^2\gg m^2,k^2$, the dynamics is that of a supercritical system, and the Universe would
eventually collapse in a ``Big Crunch" singularity.

\section{Massive Modification}
In the present section, we will describe the effects a massive modification -- of the kind first
described in \cite{Gabadadze:2004iv} -- has when it is considered in conjunction with the deformation described in
section~\ref{sect:stopping.exp}.

Before starting on describing the cosmological evolution that arises when
both deformations are taken into account, we should
emphasize that the present modification is very different from the one so far described.
We showed that the number of propagating degrees of freedom is unaltered when $\sqrt{-g}f(\sqrt\gamma)$
is introduced. Instead, in the case we shall present the number of degrees of freedom will be different.

The Lagrangian of the model we would like to discuss is
\begin{equation}
	\Lagr = \sqrt{-g}[\R-2\Lambda+2\mu_1^2\frac{(1-N)^2}{N}+2\mu_2^2\gamma^{\alpha/2}]\,,
	\label{lagr.tuttifrutti}
\end{equation}
with the following equations of motion
\begin{eqnarray}
	G_{\mu\nu}
	&+&\left[ \Lambda-\mu_1^2\frac{(1-N)^2}{N}-\mu_2^2(\alpha+1)\gamma^{\alpha/2} 
			\right]g_{\mu\nu}+\nonumber\\
	&+&\left[ \mu_1^2N(N^2-1)-\mu_2^2\alpha N^2\gamma^{\alpha/2}
			\right]\delta_\mu^0\delta_\nu^0=0\,.
	\label{eom.tuttifrutti}
\end{eqnarray}
Notation is consistent with the one used before. $N$ is the lapse function and $\gamma_{ij}$ the
induced spatial metric defined in \eqref{3D.vars}. As we discussed in our previous letter \cite{Gabadadze:2004iv}, the
modification $\sqrt{-g}(1-N)^2/N$ is the only one quadratic in the lapse function, over which fluctuations
do not present instabilities like tadpoles and the Hamiltonian of \eqref{lagr.tuttifrutti} is bounded from below.

It is evident from the Hamiltonian
\begin{equation}
	\H=\sqrt{\gamma}
	\left[ NR^0+N_jR^j+2\Lambda N-2\mu_1^2(1-N)^2-2\mu_2^2N\gamma^{\alpha/2} \right]\,,
	\label{ham.tuttifrutti}
\end{equation}
that the shift function $N_j$ remains a Lagrange multiplier, while the lapse $N$ ceases to be it.
The algebraic equation for $N$, that is
$2\mu_1^2\,N=R^0/2+\Lambda+\mu_1^2-\mu_2^2\,\gamma^{\alpha/2}$, can be seen as a constraint
for the degrees of freedom, thus on this background the gravitational field propagates three
degrees of freedom, instead of the two as in the previous cases.

For this, the newly added modification is
different from the ones we discussed previously: despite our analysis will follow closely the one
of the previous sections, we should bear in mind that the models describe two very different fields.
On the one hand, in section~\ref{sect:stopping.exp} the field is a massless tensor field propagating
on a deformed (with respect to standard GR) background; on the other, in the present section we will
discuss a model for a massive\footnote{in this context, we call ``massive'' a field that propagates
a number of degrees of freedom different than two; it should be noted also that the action is not
symmetric under the Lorentz symmetry and therefore the intuition of a massive tensor field propagating
five degrees of freedom is not necessarily respected.} tensor field.
The presence of the deformations will modify the cosmological evolution away from GR, as we will
readily see in a moment.

On the ansatz $\de s^2=-N(t)^2\de t^2+a(t)\de\vec x^2$ and after a bit of algebraic manipulation of the equations
of motion \eqref{eom.tuttifrutti}, we find the following equation for the scale factor~$a$
\begin{equation}
	\dot{a}^2-\frac{\Lambda}{3}a^2
	\left[ 1+2\frac{\mu_1^2}{\Lambda}(1-\sqrt{1+a^{-3}})-\frac{\mu_2^2}{\Lambda}a^{3\alpha} \right]=0\,,
	\label{H.tuttifrutti}
\end{equation}
where the derivative $\dot{a}$ is in respect to the proper time $d\tau=N\,dt$ and $N=\sqrt{1+a^{-3}}$,
relation that can be derived directly from the equations of motion.

It is useful to discuss the solution of \eqref{H.tuttifrutti}
as for an analogous one-dimensional point particle moving in the potential $V(a)=-\frac{\Lambda}{3}a^2
\left[ 1+2\frac{\mu_1^2}{\Lambda}(1-\sqrt{1+a^{-3}})-\frac{\mu_2^2}{\Lambda}a^{3\alpha} \right]$.

\begin{figure}[t]
	\begin{center}
	    	\epsfxsize=4in
		\epsffile{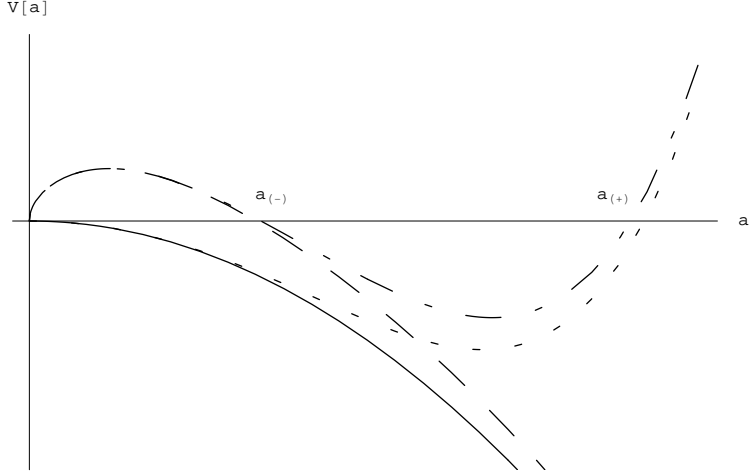}
	\end{center}
	\caption{Potential for the analogous 1-dimensional dynamics: continuous line is for $\mu_1=\mu_2=0$,
	long dashes for $\mu_2=0$, short for $\mu_1=0$ and mixed dashed line for the general case.}
	\label{fig:tuttifrutti}
\end{figure}

The potential is depicted in Fig.~\ref{fig:tuttifrutti} for various values of $\mu_1$ and $\mu_2$.
In particular, the continuous line is for $\mu_1=\mu_2=0$ and the dynamics is that of a de~Sitter
space-time with cosmological constant $\Lambda$, as expected. The other two limiting cases are when
one of the two deformations is absent; we find that for $\mu_1=0$ -- short dash curve --
the cosmological expansion proceeds as for a de~Sitter universe until a maximum size $a_{(+)}$,
after which it goes through an exponentially contracting phase. That is, the solution
we discussed previously in so much detail is recovered.

For $\mu_2=0$ -- long dash curve -- a minimum size $a_{(-)}$ for the scale factor emerges. The dynamics is very
much similar to the one we described for the cyclic universe model of section~\ref{sect:cyclic} when only
the $k^2$-deformation of \eqref{lagr.cyclic} is present: the cosmological evolution is that of a contracting
de~Sitter universe until the Universe reaches size $a_{(-)}$, after which it bounces back toward an
expanding de~Sitter phase.

When, instead, both deformations are present, the analogous potential $V(a)$ displays two turning points
$a_{(\pm)}$. The dynamics can be read easily from it, and it is like the cyclic Universe model of
section~\ref{sect:cyclic}: the cosmological evolution goes through subsequent periods of expansion and
contraction, between the minimum and maximum sizes $a_{(-)}$ and $a_{(+)}$. Those values are the zeros
of $V(a)$, and given that, when the size of the Universe is of order of $a_{(-)}$ the $\mu_2^2$-term is
sub-dominant in respect to both the cosmological constant $\Lambda$ and the $\mu_1^2$-term
(and vice-versa for $a_{(+)}$), they are approximately:
\begin{eqnarray}
	a_{(-)}&\sim&\left( \frac{4\mu_1^4}{\Lambda(\Lambda+4\mu_1^2)} \right)^\frac{1}{3}\,,\nonumber\\
	a_{(+)}&\sim&\left( \frac{\Lambda}{\mu_2^2} \right)^\frac{1}{3\alpha}\,.
	\label{minmax.tuttifrutti}
\end{eqnarray}
The turning point $a_{(+)}$ is the same as the one we found in the model of section~\ref{sect:stopping.exp}.
On the other hand, $a_{(-)}\sim(\mu_1/\Lambda)^{2/3}$, when $\mu_1^2\ll\Lambda$, is what would be the
turning point for a deformation of the kind $\gamma^{\alpha/2}$ with $\alpha=-1/2$. This is not surprising.
In the regime when $\mu_2$ is sub-dominant, the cosmic evolution is driven by the $\mu_1^2$-deformation.
From the equations of motion \eqref{H.tuttifrutti}, we find cosmology being driven by a term
-- $\sqrt{1+a^{-3}}\sim a^{-3/2}$ -- that acts like $a^{3\alpha}$ for $\alpha=-1/2$, {\it i.e.},
equivalent to the deformation of section~\ref{sect:stopping.exp} for a particular choice of $\alpha$.

Again we should stress that, even though
the backgrounds are alike, the fields propagating over them act very differently, for one carries
two degrees of freedom, while the other three. Therefore, they describe two very different gravitational
models.

\section{Conclusions}
Models, that describe gravity beyond Einstein GR, have been focussed mainly on higher derivative deformations,
like $f(\R)$ gravity for instance.
In the present letter, we discussed a class of models, that modify gravity via potential-like terms.

The $f(\sqrt{\gamma})$ deformation we introduced does not add any derivatives of the metric to the action.
Because of this, we argued that the propagating degrees of freedom are of a transverse-traceless tensor
field, as in GR. We arrived to this conclusion by studying the perturbations of the metric in the 
Lagrangian formalism. Therein, it was also evidenced the presence of instantaneous interactions. This a-casual
effect is characteristic of the models at hand, where Lorentz symmetry is broken explicitly. Their presence was
firstly noted in \cite{Gabadadze:2004iv}, and we remind to it for a more detailed discussion.

We then studied the exact cosmological solution for some particular choices of $f(\sqrt{\gamma})$. 
The main feature is that the introduced term acts to generically stop the cosmological evolution.
Depending on the details of the model, a bounce is generated during either a period of contraction or
one of expansion.
Independently from these details though, particles are produced via quantum effects at the bounce. Quantum
corrections destabilize an otherwise stable classical solution. The result is to create a future ``Big Crunch''
singularity: after a finite time the scalar curvature will diverge and the Universe will shrink to zero size.

In more specific sections, we discussed the effects and bounds on the parameters that arise when the model
is embedded into a generic model of inflation. In particular, we noticed that the dimension-full parameter $m^2$
has to be exponentially smaller than the Hubble radius during inflation, so to satisfy the bounds
imposed by the CMB.

Left to future investigations is the study of Schwarzschild solutions for the present model, like in 
\cite{Gabadadze:2005qy,Gabadadze:2007as,Dubovsky:2007zi}. It would be interesting
to know what kind of effects the $f(\sqrt{\gamma})$-term has on a Schwarzschild-like solution,
and to see how (if) it screens the gravitational field of a massive point-particle.

Also left out is the understanding of the UV completion of the model.
We have always stressed that the introduced deformation
$f(\sqrt{\gamma})$ should be thought as an effective term rising from some UV phenomena. The question is to
exactly determine what kind of phenomena. Studies in this direction have flourished in the past few years,
\cite{ArkaniHamed:2003uy} and
\cite{Ganor:2006ub}-\nocite{Ganor:2007qh,Dubovsky:2006vk,Dubovsky:2004sg,Bluhm:2007bd,Bluhm:2007gs}\cite{Bluhm:2008rf},
and it would be interesting to have a fully consistent mechanism that breaks spontaneously the Lorentz symmetry
and generates $f(\sqrt{\gamma})$-term at low energies.

\section*{Acknowledgements}

I would like to thank Gia Dvali, Gregory Gabadadze and Oriol Pujol\`as for useful discussions.
This work is supported by  Physics Department Graduate Student Fellowship at NYU.

\bibliographystyle{utphys}
\bibliography{LVdS}

\end{document}